\documentclass[a4paper,11pt]{article}
\pdfoutput=1

\usepackage[utf8]{inputenc}
\usepackage{arydshln} 
\usepackage{xcolor}
\usepackage{float}

\usepackage{epsf}

\usepackage{graphicx}

\usepackage{verbatim} 

\usepackage{amsmath}
\usepackage{amssymb}
\usepackage{mathrsfs}

\usepackage{amsfonts}
\usepackage{xfrac} 
\usepackage{booktabs} 
\usepackage{soul} 
\usepackage{pifont} 
\graphicspath{{figs/}}

\usepackage{cite}
\usepackage[
colorlinks=true
,urlcolor=black
,anchorcolor=black
,citecolor=black
,filecolor=black
,linkcolor=black
,menucolor=black
 ,pagecolor=black
,linktocpage=true
,pdfproducer=medialab
,pdfa=true
]{hyperref}

\usepackage[left=2cm,right=2cm,top=2cm,bottom=3cm]{geometry}

\usepackage{color}
\definecolor{cred}{RGB}{180,50,40} 
\definecolor{purple}{RGB}{180,90,180} 
\definecolor{darkgreen}{RGB}{0, 100, 0}

\def\be{\begin{equation}}
\def\ee{\end{equation}}
\newcommand{\bea}{\begin{eqnarray}}
\newcommand{\eea}{\end{eqnarray}}

\allowdisplaybreaks
\begin{document}

\setcounter{footnote}{0}
\vspace*{-1.5cm}
\begin{flushright}
LPT Orsay-19-17 \\

\vspace*{2mm}
\end{flushright}
\begin{center}
\vspace*{1mm}

\vspace{1cm}
{\Large\bf 
Interference effects in LNV and LFV semileptonic decays: \\}

\vspace*{0.2cm}{\Large\bf the
  Majorana hypothesis}

\vspace*{0.8cm}

{\bf A.~Abada$^{a}$, C.~Hati$^{b}$, X.~Marcano$^{a}$ and A.~M.~Teixeira$^{b}$}  
   
\vspace*{.5cm} 
$^{a}$ Laboratoire de Physique Th\'eorique, CNRS\\
Univ. Paris-Sud, Universit\'e Paris-Saclay, 91405 Orsay, France

\vspace*{.2cm} 
$^{b}$ Laboratoire de Physique de Clermont (UMR 6533), CNRS/IN2P3,\\ 
Univ. Clermont Auvergne, 4 Av. Blaise Pascal, F-63178 Aubi\`ere Cedex,
France 
\end{center}

\vspace*{2mm}

\begin{abstract}
In the case  where the Standard Model is extended by one heavy
Majorana fermion, the  branching fractions of semileptonic meson decays  into 
same-sign and opposite-sign dileptons are expected to be of the same
order. As we discuss here, this need not be the case in
extensions by at least two sterile fermions, due to the
possible destructive and constructive interferences that might arise.
Depending on the  $CP$ violating phases, one can have
an enhancement of the lepton number violating 
modes and suppression of the lepton number conserving ones (and
vice-versa). We explore for the first time the interference effects
in semileptonic decays, and 
illustrate them for  a future observation of kaon decays at NA62.
We also argue that a non-observation of a given mode need not be
interpreted in terms of reduced active-sterile mixings, but that it
could instead be understood in terms of interference effects due to
the presence of several sterile states; 
in particular, for different-flavour final state charged leptons, 
observing a lepton number conserving process and not a
lepton number violating one does 
not rule out that the mediators are Majorana fermions. 
\end{abstract}
\vspace*{1mm}

\section{Introduction}

Several extensions of the Standard Model (SM) aiming at explaining
oscillation phenomena call upon the introduction of right-handed
(RH) neutrinos, and the embedding of the seesaw mechanism onto the SM
is one of the most economical mechanisms for the generation of
neutrino masses and lepton mixings.   
The presence of relatively light RH neutrinos (sterile
fermions from the gauge point of view), which  
have non-negligible mixings with the active ones, leads 
to the modification of the charged and neutral lepton currents, and thus
the new states can give rise to  
contributions to many observables, ranging from rare transitions and
decays at high-intensities to signals at colliders. 
Mechanisms of
neutrino mass generation relying on such sterile fermions are thus
particularly appealing, as they offer the possibility of being (albeit
indirectly) tested.

The observation of a lepton number violating (LNV) process will
undoubtedly point towards the existence of New Physics (NP) and
indirectly to the clear Majorana nature of the exchanged fermion. 
Lepton number violating meson and tau semileptonic decays
are examples of such observables; currently, they are the object of
extensive world-wide searches, for instance of NA62 (for the light
mesons), BES-III for charmed mesons, and LHCb and Belle~II for
B-mesons and tau leptons. 

In this work we address the LNV  
meson and tau lepton semileptonic three-body decays, 
$M_1^\pm\to \ell_\alpha^\pm \ell_\beta^\pm M_2^\mp$, and $\tau^\pm \to
\ell_\alpha^\mp M_1^\pm M_2^\pm$, as well as their 
lepton number conserving (LNC) counterparts, 
$M_1^\pm\to \ell_\alpha^\pm\ell_\beta^\mp M_2^\pm$, 
and $\tau^\pm \to \ell_\alpha^\pm M_1^\pm M_2^\mp$ (with the different
mesonic states $M_{1,2}$ 
and charged lepton flavours $\alpha$ and $\beta$ 
including all kinematically allowed possiblities).
When mediated by sterile fermions, the joint study of the LNC and LNV
processes allows to shed light on the nature (Dirac or Majorana)
of the exchanged mediator.

In the presence of Majorana heavy neutral leptons 
(a right-handed neutrino, or other sterile fermions), one can 
have sizeable branching fractions (BRs) for lepton number
violating processes, when the Majorana states are
produced on-shell~\cite{Flanz:1999ah,Zuber:2000vy,Atre:2005eb,
Atre:2009rg,Rodejohann:2011mu,Helo:2011yg,Liu:2016oph,Quintero:2016iwi,
Zamora-Saa:2016qlk,Zamora-Saa:2016ito,Abada:2017jjx}
 - the so-called ``resonant production'' regime. 
In the case in which a single sterile fermion is present, 
the branching fractions for the semileptonic decays leading to 
same-sign and opposite-sign dileptons are predicted to be of the same
order. 
Notice that due to the
existence of a single heavy fermion, the decay widths are not sensitive
to the latter's $CP$ Majorana phases - in particular, one can still
have contributions to LNV decays even if all Majorana $CP$ phases 
are set to zero.
Interestingly, should more than one heavy Majorana state be present, 
the relative size of the LNV and LNC channels may be very different due to
interference effects associated with the Dirac and Majorana phases.  

In this work, we consider the SM extended by two sterile fermion
states which mix with the three active neutrinos; we take the 
neutrino mass eigenvalues and the lepton mixing matrix to be
independent, implying that no assumption is made on the mechanism of
neutrino mass generation. 
Other than being in the mass range to be produced ``on-shell'' in the
semileptonic decays, the two new states are taken to be sufficiently
close in mass to allow for sizeable interference effects
(constructive or destructive) in the LNV and LNC decays, 
in association with the new $CP$-violating phases.  

As we will discuss in detail, depending on the sterile fermion
parameter space (i.e., their masses,  mass difference and 
mixings to the active neutrinos - including the  $CP$ violating 
phases), the interference effects can lead to various scenarios, 
ranging from an extreme suppression of the 
lepton number conserving observables
while having large contributions to lepton number violating ones (and
this contrary to the usual belief), to 
the reverse case. Our findings, relying on a complete numerical analysis,  
strongly suggest that effects leading to suppressions 
in some channels should be taken into account upon 
the interpretation of a non-observation of LNV or LNC meson (or tau) 
semileptonic decays:
reduced BRs might result from the interference of (at least) two
sterile states with small mass difference(s), and should not be
necessarily translated 
into more stringent bounds on the heavy neutrino couplings to the active
ones. 
More importantly, the observation of a lepton number conserving 
process accompanied by negative search results for the corresponding LNV modes 
(with both predicted to be within experimental sensitivity)  
does not necessarily imply that the mediators are of Dirac nature. 

Here  we present the results of our study, focusing on the 
semileptonic kaon decays currently searched for in NA62, that is 
$K^\pm \to\ell_\alpha^\pm\ell_\beta^\pm \pi^\mp$ decays and the
corresponding lepton number conserving ones,  $K^\pm
\to\ell_\alpha^\pm\ell_\beta^\mp \pi^\pm$ (where $\alpha,\beta$ 
denote electrons and/or muons). Kaon decays are used here as an
illustrative example, as the results generically hold 
for all similar semileptonic decays (including tau decays). 

Similar studies have focused on the 
prospects of LNV searches at colliders, see~\cite{Cai:2017mow}, and
references therein. 
Some works have explored the role of the second heavy neutrino
concerning the possibility of resonant $CP$
violation~\cite{Bray:2007ru}, while others have  
compared the expected number of events associated with 
same-sign and opposite-sign 
dileptons at colliders in the framework of Left-Right symmetric 
models~\cite{Das:2017hmg}.
The latter analysis considered scenarios in which the relative 
$CP$ violating phases of active-sterile mixings were 
identical for the heavy neutrinos. In our work - aiming at studying
rare meson decays at lower energies - 
we relax this restrictive hypothesis,  
which opens the possibility of very distinct behaviours for the LNV 
and LNC rates, due to the interference effects.

Although the study is carried for a simplified SM extension (the
``3+2'' minimal model), we emphasise that our results generically
apply to complete mechanisms of neutrino mass generation, provided at least
two additional states are present, as in the case of 
low-scale type I seesaw frameworks (in which at least two RH neutrinos 
are required to accommodate oscillation data).

This work is organised as follows: after an analytical 
discussion in Section~\ref{mesondecays}
of semileptonic meson decay LNC and LNV amplitudes (in
the presence of two additional sterile states), 
we explore in Section~\ref{results} the
interference effects in a generic way, identifying critical regimes
and the potential consequences for the relative size of the BRs. 
Section~\ref{numerical:kaon} is dedicated to 
a full numerical analysis, in which the intereference
effects are illustrated for kaon decays.
We discuss further points, and 
summarise our most important findings in the Conclusions.

\section{Semileptonic meson decays with two sterile
  neutrinos}\label{mesondecays}  
As stated in the Introduction, we work in the framework of simplified
SM extensions via the addition of $N$ extra neutral Majorana fermions, 
making no assumption on the mechanism of
neutrino mass generation (i.e., considering 
neutrino masses and lepton mixings to be
independent). 
In the presence of new sterile states with non-negligible mixings
to the (light) active neutrinos, the leptonic charged current is modified as
\begin{equation}\label{eq:cc-lag}
- \mathcal{L}_\text{cc} \,=\, \frac{g}{\sqrt{2}} \,U_{\alpha i}\, 
\bar{\ell}_\alpha \,\gamma^\mu \,P_L \,\nu_i \, W_\mu^- + \, \text{H.c.}\,,
\end{equation}
in which $i$ denotes the physical neutrino states, from 1 to $3+N$,
and $\alpha$ the flavour of the charged leptons. 
For the case $N=2$ (corresponding to the addition of two states
with masses $m_{4,5}$), 
the unitary matrix $U$, which encodes flavour mixing in charged current
interactions, can be parametrised in terms of 
ten rotation matrices and 4 Majorana phases as follows~\cite{Abada:2016awd,Abada:2018qok}
\begin{equation}\label{Uparam}
U\,=\,R_{45}\,R_{35}\,R_{25}\,R_{15}\,R_{34}\,R_{24}\,R_{14}\,
R_{23}\,R_{13}\,R_{12}\,
\mathrm{diag}\left(1,\,e^{i\varphi_2},\,e^{i\varphi_3},\,e^{i\varphi_4},\,
e^{i\varphi_5}\right)\,.
\end{equation} 
In the above, 
$R_{ij}$ corresponds to the rotation matrix between the $i$ and $j$ states
(each parametrised by a mixing angle $\theta_{ij}$ and a Dirac
$CP$-violating phase $\delta_{ij}$) and
$\varphi_i$ represent Majorana $CP$-violating phases.
For instance, the rotation matrix $R_{45}$ can be explicitly cast as  
\begin{equation}
R_{45}\,=\,\left(
\begin{array}{ccccc}
1 & 0 & 0 & 0 & 0\\
0 & 1 & 0 & 0 & 0\\
0 & 0 & 1 & 0 & 0\\
0 & 0 & 0 & \cos\theta_{45} & \sin\theta_{45}\,e^{-i\delta_{45}}\\
0 & 0 & 0 & -\sin\theta_{45}\,e^{i\delta_{45}} & \cos\theta_{45}\\
\end{array}
\right),
\end{equation}
and similarly for the other $R_{ij}$. 
Since several of the Dirac phases are non-physical\footnote{Note that
  in the case of a $3+N$ model,  
the mixing matrix $U$ includes a total of 
$(3+N)(2+N)/2$ rotation angles, 
$(2+N)(1+N)/2$ Dirac phases and $2+N$ Majorana phases. Together with 
the masses of the new sterile states, 
$m_i$, $i=1,\dots, N$, the latter
constitute the physical parameters of the model.}, 
we thus set
$\delta_{12}=\delta_{23}=\delta_{24}=\delta_{45}=0$.  The
parametrisation of Eq.~\eqref{Uparam}, which  ensures the unitarity of
the full mixing matrix, allows to clearly single out the nature of the
$CP$ phases (Dirac or Majorana). The mixing between the left-handed
leptons corresponds to a $3 \times 3$ block of $U$, which is
non-unitary due to the new mixings
with the heavy neutrinos. 

In the following we will be interested in the mixings of the sterile
states  to the active sector. Using the parametrisation of 
Eq.~\eqref{Uparam}, these can be written as 
\begin{equation}\label{active-sterile}
\left(\begin{array}{cc}
U_{e4} & U_{e5} \\
U_{\mu4} & U_{\mu5}\\
U_{\tau4} & U_{\tau5}
\end{array}\right)
\approx
\left(\begin{array}{cc}
s_{14} e^{-i(\delta_{41}-\varphi_{4})} & s_{15} e^{-i(\delta_{51}-\varphi_{5})} \\
s_{24} e^{i\varphi_{4}} & s_{25} e^{-i(\delta_{52} -\varphi_{5})} \\
s_{34} e^{-i(\delta_{43}-\varphi_{4})} & s_{35} e^{-i(\delta_{53}-\varphi_{5})}
\end{array}\right)\,,
\end{equation}
where $s_{ij}=\sin\theta_{ij}$ and where we have neglected terms of
$\mathcal O(s_{ij}^2)$.  \\
\noindent
We denote the active-sterile mixing elements by 
\be\label{angle-phase}
U_{\alpha i}=e^{-i\phi_{\alpha i}}|U_{\alpha i}|, \quad
\alpha=e,\mu,\tau, \text{ and }  i=4,5\ ,  
\ee
where each $\phi_{\alpha i}$ is a combination of the 7 $CP$-violating
phases (5 Dirac and 2 Majorana) in Eq.~\eqref{active-sterile}.
We notice that in the framework of this 
simplified model one  can,  without any loss of generality,
choose the mixing angles  and the (Dirac and Majorana)  phases  as
independent input parameters.  

\bigskip
We now address 
the effect of the new mixings on the LNC  semileptonic
processes $M\to M' \ell_\alpha^\pm \ell_\beta^\mp$  and the
corresponding LNV ones $M\to M' \ell_\alpha^\pm \ell_\beta^\pm$, $M$
and $M'$ being pseudoscalar mesons\footnote{Here we present the case of
  semileptonic meson decays; however, a similar discussion holds for
  semileptonic tau decays $\tau \to M \, M^\prime \ell_\alpha$,
  $\alpha = e, \mu$. Moreover, for simplicity we
  focus on $M^+\to M^{'\!+} \ell_\alpha^+ \ell_\beta^-$ and $M^+\to
  M^{'\!-} \ell_\alpha^+ \ell_\beta^+$.}. 
Their  squared amplitudes 
(see~\cite{Abada:2017jjx}) are proportional, up to overall constant
parameters, to the following: 

\begin{align}
\left|\mathcal{A}^{\rm LNC}_{M\to M' \ell_\alpha^+ \ell_\beta^-}\right|^2 \!\!&\propto
\Big| U_{\alpha 4} U^*_{\beta 4} g(m_4) + U_{\alpha 5} U^*_{\beta 5}
g(m_5) \Big|^2 ,\!\!\! \quad \left|\mathcal{A}^{\rm LNC}_{M\to M'
  \ell_\alpha^- \ell_\beta^+}\right|^2 \!\!\propto 
\Big| U^*_{\alpha 4} U_{\beta 4} g(m_4) + U^*_{\alpha 5} U_{\beta 5}
g(m_5) \Big|^2\nonumber\\
  \left|\mathcal{A}^{\rm LNV}_{M\to M' \ell_\alpha^+ \ell_\beta^+}\right|^2 \!\!&\propto
\Big| U_{\alpha 4} U_{\beta 4} f(m_4) + U_{\alpha 5}
U_{\beta 5} f(m_5) \Big|^2 \!,\!\!\! \quad  \left|\mathcal{A}^{\rm
  LNV}_{M\to M' \ell_\alpha^- \ell_\beta^-}\right|^2 \!\!\propto 
\Big| U^*_{\alpha 4} U^*_{\beta 4} f(m_4) + U^*_{\alpha 5}
U^*_{\beta 5} f(m_5) \Big|^2 \!\nonumber
\end{align}
leading to 
\begin{align}
\left|\mathcal{A}^{\rm LNC}_{M\to M' \ell_\alpha^\pm
  \ell_\beta^\mp}\right|^2 \!\!&\propto 
 \big|U_{\alpha 4}\big|^2 \big|U_{\beta4}\big|^2
|g(M)|^2\, \Big|1+\kappa'\, e^{\mp i(\psi_{\alpha
  }-\psi_{\beta})}\Big|^2\,,\label{GamamLNC}\\ 
\left|\mathcal{A}^{\rm LNV}_{M\to M' \ell_\alpha^\pm
  \ell_\beta^\pm}\right|^2 \!\!&\propto 
 \!\big|U_{\alpha 4}\big|^2
\big|U_{\beta4}\big|^2 |f(M)|^2\, \Big|1+\kappa\, e^{\mp
  i(\psi_{\alpha}+\psi_{\beta})}\Big|^2 , \label{GamamLNV} 
\end{align}
where we have defined  $\psi_\alpha\equiv
\phi_{\alpha_5}-\phi_{\alpha4}$, and $M$ is the average mass of the
two sterile neutrinos ($\Delta M$  their mass splitting), so that
$m_4=M-\Delta M/2$ and $m_5=M+\Delta M/2$. 
The functions  $f$ and $g$ are the integrals one obtains when
computing the decay amplitudes for LNV and LNC semileptonic decays of
mesons  (details can be found in for instance \cite{Abada:2017jjx}).
The complex quantities $\kappa$ and $\kappa'$, which reflect the
relative size of the contributions of the two sterile fermions to the
processes, are defined as 
\begin{equation}\label{kappas}
\kappa\equiv\dfrac{ |U_{\alpha5} U_{\beta5}| }{|U_{\alpha4}
  U_{\beta4}|}\dfrac{f(m_5)}{f(m_4)}, \quad \kappa'\equiv\dfrac{
  |U_{\alpha5} U^*_{\beta5}| }{|U_{\alpha4}
  U^*_{\beta4}|}\dfrac{g(m_5)}{g(m_4)} \,.
\end{equation} 

Equations~(\ref{GamamLNC},~\ref{GamamLNV}) allow to infer several
important points:  
as expected, the LNC decay amplitudes are not sensitive to the Majorana
$CP$ violating phases $\varphi_i$, as these cancel out in the
$\psi_{\alpha}-\psi_{\beta}$ combination; the LNC decay amplitudes are 
sensitive to the Dirac phases, but only in the case of flavour violating final
states, i.e.  $\alpha\ne\beta$~\cite{Bray:2007ru}.   
On the other hand, the LNV decay amplitudes are sensitive to both  Dirac
and Majorana $CP$ phases (since the phase appearing in the decay amplitude
is the sum of the relative $CP$ phases, $\psi_{\alpha}+\psi_{\beta}$),
and this even in the case of identical charged leptons in the final
state ($\alpha=\beta$).  

\bigskip
In order to discuss the impact of the $CP$ phases on the LNV and LNC
decay amplitudes, as well as possible interference effects, we
consider the quantity $R_{\ell_\alpha\ell_\beta} $ defined as 
\begin{equation}\label{ratio:simple}
R_{\ell_\alpha\ell_\beta} \equiv
\frac{\Gamma^{\rm LNV}_{M\to M' \ell_\alpha^\pm
    \ell_\beta^\pm}}{\Gamma^{\rm LNC}_{M\to M'
    \ell_\alpha^\pm \ell_\beta^\mp}} \,, 
\end{equation}
and further introduce  the ratio $\widetilde R_{\ell_\alpha\ell_\beta}$ 
\begin{equation}\label{Rtilde}
\widetilde R_{\ell_\alpha\ell_\beta} \equiv
\frac{\Gamma^{\rm LNC}_{M\to M' \ell_\alpha^\pm
    \ell_\beta^\mp}-\Gamma^{\rm LNV}_{M\to M'
    \ell_\alpha^\pm \ell_\beta^\pm}} 
{\Gamma^{\rm LNC}_{M\to M' \ell_\alpha^\pm
    \ell_\beta^\mp}+\Gamma^{\rm LNV}_{M\to M'
    \ell_\alpha^\pm \ell_\beta^\pm}} 
=\frac{1-R_{\ell_\alpha \ell_\beta}}{1+R_{\ell_\alpha \ell_\beta}}\,,
\end{equation}
with, in both ratios, $\Gamma^{\rm LNC}_{M\to M' \ell_\alpha^\pm
  \ell_\beta^\mp} \equiv \Gamma^{\rm LNC}_{M\to M'
  \ell_\alpha^+ \ell_\beta^-} +\Gamma^{\rm LNC}_{M\to M'
  \ell_\alpha^- \ell_\beta^+}$, in the case in which $\alpha\ne
\beta$. 

\noindent
The ratio $R_{\ell_\alpha\ell_\beta} $ is  usually
considered to compare  LNV to LNC processes 
(a similar approach to what was
done, for instance, in the context of collider 
searches~\cite{Das:2017hmg}) and  
the second ratio, $\widetilde R_{\ell_\alpha\ell_\beta}$, which a priori
might seem redundant, will be useful to better understand interference
effects.

\section{Exploring the interference effect}
\label{results}

As extensively
discussed in~\cite{Flanz:1999ah,Zuber:2000vy,Atre:2005eb,
Atre:2009rg,Rodejohann:2011mu,Helo:2011yg,Liu:2016oph,Quintero:2016iwi,
Zamora-Saa:2016qlk,Zamora-Saa:2016ito,Abada:2017jjx}, in addition to
being of Majorana nature,  
the sterile fermions mediating the LNV decays should be 
 produced on-shell, in which case  one can have a resonant
enhancement of the decay widths. In the narrow-width 
approximation, this  ``resonant enhancement'' can be understood as an
increase of $\mathcal O(m_i/\Gamma_{N_i})$ in the decay rates
($\Gamma_{N_i}$ denoting the width of the heavy sterile state $N_i$). For
this reason, we will assume  
 that the individual widths are very small compared to the sterile
 neutrino masses\footnote{Notice that this assumption is well
   justified, as this is usually the case in seesaw-like models where
   the sterile neutrinos are lighter than the typical meson 
   masses~\cite{Bondarenko:2018ptm}.},
 $\Gamma_{N_i}\ll m_i$.

In the case of the SM extended by only one  heavy Majorana
neutrino, we have verified that 
the predictions for the LNV and LNC decay widths are of the
same order, implying that $R_{\ell_\alpha \ell_\beta}=1$ and thus
$\widetilde R_{\ell_\alpha\ell_\beta} =0$. 
In the presence of two (or more) sterile fermions with
(clearly) non-degenerate masses, interference effects are negligible
and one recovers the previous predictions for $R$ and $\widetilde R$.  
However, when the mass splitting of the heavy Majorana states is very
small, one can  have an overlap  between their  contributions,
possibly leading to destructive or constructive interferences. The
effect of the overlap will be maximal should  the mass splitting be
even smaller than the Majorana neutrino decay widths.  In turn, this
will lead to different predictions for the LNV and LNC decay widths,
changing the values $R$ and $\widetilde R$.  
In summary, interference effects are expected to be  relevant if both
the following conditions are realised:  
\begin{equation}\label{conditions}
\Delta M\ll M \quad\text{and}\quad \Delta M < \Gamma_N\,,
\end{equation}
in which, for simplicity, we have assumed the widths to be the same
$\Gamma _{N_4}=\Gamma_{N_5}= \Gamma_N$. 
With these conditions, and in terms of the $CP$-violating phases, 
the ratio $R_{\ell_\alpha\ell_\beta}$ is  given as follows 
\begin{equation}\label{ratio:simple-gen}
R_{\ell_\alpha\ell_\beta}=
\frac{(1- |\kappa|)^2 + 4 |\kappa|
  \cos^2\left({\delta\pm(\psi_\alpha+\psi_\beta)\over 2}\right)}{(1-
  |\kappa'|)^2 + 4 |\kappa'|
  \cos^2\left({\delta'\pm(\psi_\alpha-\psi_\beta)\over 2}\right)} \,, 
\end{equation}
where we have set $\kappa^{(')}= |\kappa^{(')}|e^{ i \delta
  ^{(')}}$, and with the $\pm$ referring to the electric charge of the
lepton $\alpha$. 

\noindent Moreover,  
the coefficients $\kappa$ and $\kappa'$ of Eq.~(\ref{kappas}) can be
expanded as follows 
\begin{equation}\label{kappa}
|\kappa|\simeq |\kappa'|
= \dfrac{ |U_{\alpha5} U^*_{\beta5}| }{|U_{\alpha4} U^*_{\beta4}|}
\Big(1+{\mathcal{O}} \big( \frac{\Delta M}{\Gamma_N} \big)\Big)\ .
\end{equation} 
In order to have sizeable interference effects,  in
addition to  having a small mass splitting, the relative size of
the contributions of the two neutrinos to each amplitude should be of
the same order, and not very different from 1,
$|\kappa|\sim|\kappa'|\approx 1$ (as can be 
seen from Eqs.~(\ref{GamamLNC}, \ref{GamamLNV})), implying that 
the two neutrinos should
mix with similar strength to the relevant active flavours.  

Under the hypotheses of Eq.~(\ref{conditions}), and in the limit
$|\kappa|\sim|\kappa'|\sim 1$,  one can derive the ratios $
R_{\ell_\alpha\ell_\beta}$ and  
$\widetilde R_{\ell_\alpha\ell_\beta}$  in terms of the $CP$-violating phases as 
\bea\label{ratios-cp}
R_{\ell_\alpha\ell_\beta} &=&
 \frac{\cos^2\left[{1\over
       2}(\psi_\alpha+
\psi_\beta)\right]}{\cos^2\left[{1\over 
       2}(\psi_\alpha-
\psi_\beta)\right]}\,,\\ 
\widetilde R_{\ell_\alpha\ell_\beta}
&=& \frac
           {\sin \psi_\alpha\sin\psi_\beta}{
\cos\psi_\alpha\cos \psi_\beta \ + 1}\ , 
\eea
where (for simplicity) we have assumed  
in the last equations that $\delta=\delta'=0$.
One can immediately notice from Eq.~(\ref{ratios-cp}) that, for
$\alpha\ne \beta$,  the ratio  
$R_{\ell_\alpha\ell_\beta}$ can deviate from 1 (larger or smaller) due
to the presence of 
both relative $CP$ violating phases, $\psi_\alpha$ and $\psi_\beta$.  

The effect of the interference between the two sterile fermion
contributions can already be seen in the simple limiting case in which
the relative $CP$-violating phases are identical 
 $\psi_\alpha=\psi_\beta$ (the same limit was also used
in~\cite{Das:2017hmg} regarding collider searches).  
This  situation can be realised  
if, for example, one sets all the Dirac $CP$ phases to zero so that
the ratio $R_{\ell_\alpha\ell_\beta} $ depends only on the Majorana
$CP$ phases.  
It is important to notice that in such cases (i.e., for
$\psi_\alpha=\psi_\beta$) no interference effects (destructive or constructive)
occur for the LNC case.  

\begin{figure}[t!]
\begin{center}
\includegraphics[width=.5\textwidth]{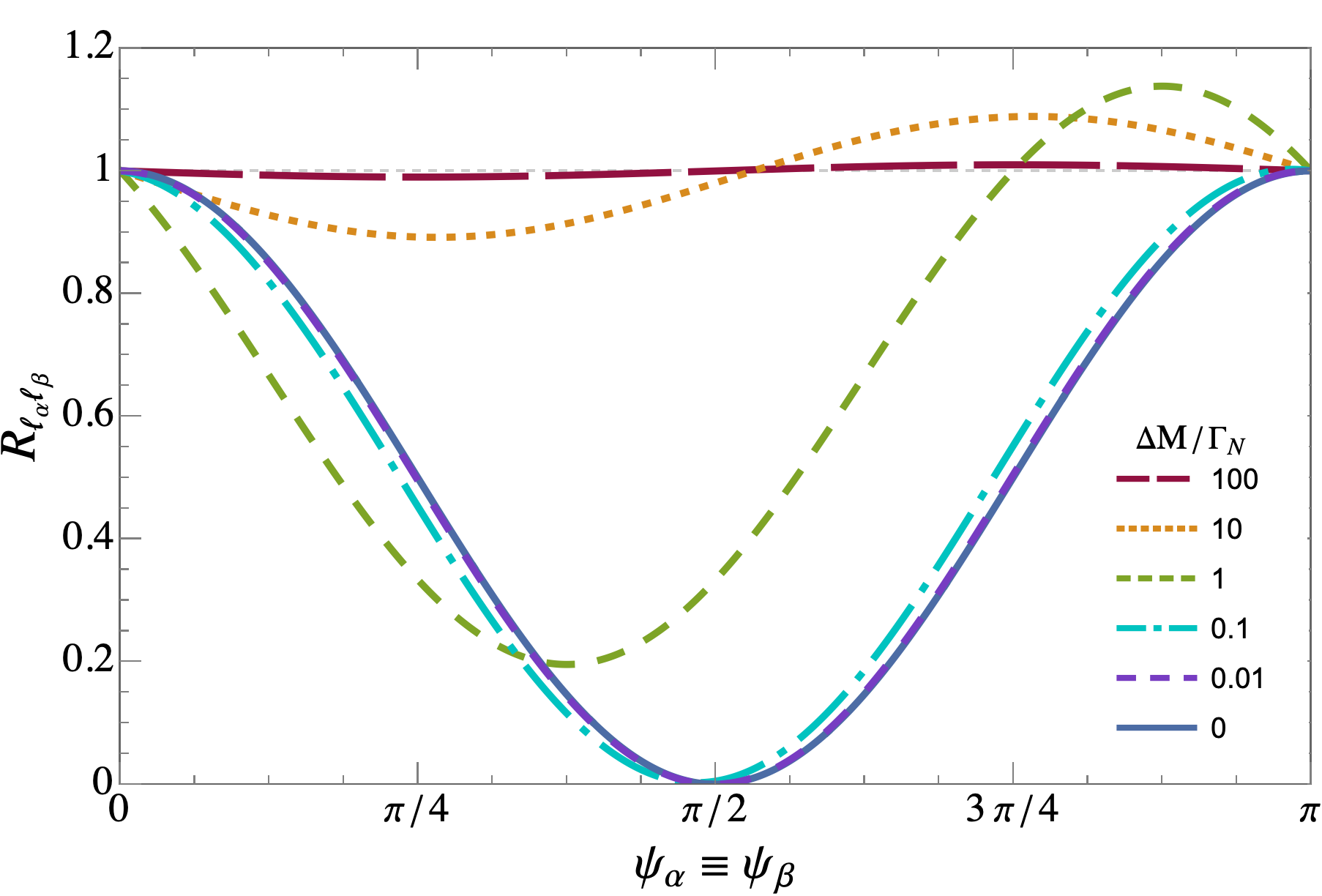}
\caption{The ratio $R_{\ell_\alpha\ell_\beta}$, as generically arising
  from the comparison of the LNC and LNV widths of any given 
  semileptonic meson decay;
  $R_{\ell_\alpha\ell_\beta}$ is depicted as a function of a common
  $CP$ violating phase, $\psi_\alpha=\psi_\beta$. Several
  regimes of $\Delta M/\Gamma_N$ illustrate how the conditions of 
  Eq.~(\ref{conditions}) are crucial to observe maximal interference
  effects.     
}\label{plotRonephase}
\end{center}
\end{figure}

In this limiting case, the ratio $R_{\ell_\alpha\ell_\beta}$
is illustrated in Fig.~\ref{plotRonephase} as a function of the common
relative $CP$ phase ($\psi_\alpha=\psi_\beta$) 
for different values of $\Delta M/\Gamma_N$.
The lines of Fig.~\ref{plotRonephase}
were obtained via a full numerical\footnote{The numerical analysis was
  done following the full computation of~\cite{Abada:2017jjx}, 
  adapted to the case
  of two additional sterile states.
} evaluation of Eq.~(\ref{ratio:simple}), and for the example of $K
\to \pi e \mu $ 
LNV and LNC decays; here we have set $M=350$~MeV, and 
taken an identical strength for the
active-sterile mixings,  $|U_{\alpha 4}| \approx |U_{\alpha 5}| =
10^{-5}$ ($\alpha = e, \mu$)\footnote{Such a choice is inspired by the
  study of~\cite{Abada:2017jjx}, since it leads to contributions to
  the LNV (and LNC) BRs within future experimental sensitivity, some
  modes being already in conflict with curent data. 
};  
the sterile neutrino total decay widths are computed
following~\cite{Atre:2009rg,Abada:2017jjx,Bondarenko:2018ptm}.\\ 
Similar analyses were carried for other
meson decays and different flavour content of the final lepton
states (when kinematically allowed); we have confirmed that in 
{\it all} cases the
corresponding $R_{\ell_\alpha\ell_\beta}$ exhibit an analogous
behaviour to that of Fig.~\ref{plotRonephase}. 
In addition, such behaviour is periodic in the phases (mod $\pi$).

One can also see from Fig.~\ref{plotRonephase}
that the ``na\"ive'' limit, $R_{\ell_\alpha\ell_\beta}=1$, is indeed
recovered for the case of vanishing values of the relative phase 
$\psi$, as expected from Eq.~(\ref{ratios-cp}).
With the exception of regimes leading
to $\Delta M/\Gamma_N \gg 1$ -- for which there is no visible
interference effect and one recovers the expectations of the SM
extended by one sterile state
(i.e., $R_{\ell_\alpha\ell_\beta} \simeq 1$)
-- the lines of Fig.~\ref{plotRonephase} illustrate
both constructive and destructive interference effects in the LNV
amplitude, clearly signalling the presence of at least
two additional sterile states. 

The constructive interference effects are typically more
important for values of $\psi \sim \pi$;  for the
curves corresponding to very small $\Delta M/\Gamma_N \ll 1$, the
constructive interference is still present (provided 
$\Delta M \neq 0$), although not visible due to 
the resolution of the plot. 

The most remarkable effect in Fig.~\ref{plotRonephase} is the 
destructive interference in the LNV amplitude due to the presence of
two sterile neutrinos, which coherently interfere:   
suppressions of order $90\%$ or above can occur when 
$\Delta M/\Gamma_N \to 0$. In this situation the interpretation of 
negative LNV searches does not preclude the observation of a signal
for the corresponding LNC channel; especially in the framework of
models containing additional Majorana fermions, this might only
suggest the presence of at least two states, with non-negligible
relative $CP$ violating phases.

Notice that in the case $\psi_\alpha=\psi_\beta$, 
our analysis of semileptonic decays - carried in a very
generic extension of the SM by two heavy neutral leptons - leads to
patterns similar to those of Ref.~\cite{Das:2017hmg} (in which the
quantity $R_{\ell \ell}$ denoted the ratio of same-sign to 
opposite-sign dilepton number of events in hadron colliders). 

\bigskip
The constructive/destructive interference effects illustrated in 
Fig.~\ref{plotRonephase} are a general feature of semileptonic meson
and tau decays. These effects can become even more pronounced if one 
steps away from the special case where the two relative $CP$ phases
are equal (which preserved LNC processes from interferences).  
In the more general case where $\psi_\alpha\ne\psi_\beta$,
interferences can occur in both LNV and LNC decay amplitudes, leading to  
enhanced values of $R_{\ell_\alpha\ell_\beta}$ - or
more strikingly, to the suppression of the LNC amplitudes. 
However, we stress that for LNC processes, interference effects can only 
occur for different flavour final state charged leptons, i.e.
$\ell_\alpha\ne \ell_\beta$. 

\noindent 
 Although we do not discuss them here, LNC and LNV four-body semileptonic
  meson decays (leading to two mesons and two charged leptons in the
  final state) could also be used to illustrate the interference effects
  due to the presence of at least two sterile states. We expect that
  the results for (a generalized definition of) $R_{\ell_\alpha
    \ell_\beta}$ would also hold.

\section{Illustrative case: semileptonic kaon decays}\label{numerical:kaon}

The suppression/enhancement of the LNV and LNC decay amplitudes
discussed in the previous section can occur in all meson (and tau)
decays  for the different (kinematically 
allowed) final state charged leptons.

An important goal of this study, previously highlighted
is the (re-)inter\-pre\-tation of possible deviations of the 
$R_{\ell_\alpha\ell_\beta}$ ratio from 1, under the working hypothesis
of the SM extended by at least two sterile fermions. In order to do so, 
one must be able to single out the heavy Majorana neutrino
contributions to both LNV and LNC decays; while 
LNV transitions are forbidden in the SM, the latter 
is the source of the dominant contributions concerning 
LNC semileptonic decays into same-flavoured lepton final states. 
From here on, we thus focus on LNC different flavour final states 
($\ell_\alpha \neq \ell_\beta$), to isolate the sterile
neutrino contributions\footnote{One could also consider the flavour
  diagonal channels, and subtract the SM contributions.}.  
Moreover, it is clear that 
any discussion of $R_{\ell_\alpha\ell_\beta}$
would ideally require an observation of the branching ratios of both
LNV and LNC 
transitions. 

Analyses of (lepton number violating and conserving)
semileptonic meson decays and tau-lepton decays 
have been conducted in the framework of the SM extended by one sterile
fermion~\cite{Flanz:1999ah,Zuber:2000vy,Atre:2005eb,
Atre:2009rg,Rodejohann:2011mu,Helo:2011yg,Liu:2016oph,Quintero:2016iwi,
Zamora-Saa:2016qlk,Zamora-Saa:2016ito,Abada:2017jjx},  and have allowed to
constrain the sterile fermion parameter space.
The recent study of~\cite{Abada:2017jjx}, taking into account all
available experimental limits, updated the existing bounds for the 
active-sterile mixings ($|U_{\ell 4}|$ and sterile
fermion mass); the results of this 
update suggest that in certain cases, in particular for semileptonic kaon
decays, the BRs of the LNV modes $K^\pm
\to\ell_\alpha^\pm\ell_\beta^\pm \pi^\mp$ ($\alpha, \beta = e, \mu$)
and of the SM forbidden LNC mode, $K^\pm \to e^\pm \mu^\mp \pi^\pm$, 
can both be within the future experimental sensitivity 
of NA62.
For the above reasons, we thus focus on the example of 
kaon semileptonic LNV and LNC decays 
to discuss the interference effects.
We extend the study of~\cite{Abada:2017jjx} (in particular,
concerning the resonant production and the narrow width approximation for
the Majorana mediator), and now consider the SM extended by 2 
Majorana fermions with masses 
$m_i\in[140, 493]$ MeV, in the conditions of Eq.~(\ref{conditions}). 
Our goal is to fully explore the
impact of the relative $CP$ violating phases as a source of
constructive or destructive interference, leading to deviations of the
LNC and LNV decay BRs from what is expected in the 
most minimal SM extension (via
only one sterile state).

The current bounds for the kaon LNV and LNC semileptonic decays are 
(see, e.g.~\cite{Tanabashi:2018oca}), 
\begin{align}\label{bounds:current}
&\text{BR}(K^+ \to \pi^+ e^- \mu^+) \leq 1.3 \times 10^{-11}\,, \quad 
\text{BR}(K^+ \to \pi^+ e^+ \mu^-) \leq 5.2 \times 10^{-10}\,;\\
&\text{BR}(K^+ \to \pi^- e^+ e^+) \leq 6.4\times10^{-10}\,,  
\quad
\text{BR}(K^+ \to \pi^- \mu^+ \mu^+) \leq 8.6\times10^{-11}\,, \\
&\text{BR}(K^+ \to \pi^- e^+ \mu^+) \leq
5.0\times10^{-10}\,,\label{bounds:current2} 
\end{align}
while the future NA62 sensitivity for the processes here
discussed is expected to be~\cite{Ceccucci:2015wpa}
\begin{eqnarray}\label{NA62:future}
\text{BR}(K^+ \to \pi^+ e^\pm \mu^\mp) \leq 0.7 \times 10^{-12}\,, 
\quad \text{BR}(K^+ \to \pi^- e^+ \mu^+) \leq 0.7 \times 10^{-12}\,, \\
\text{BR}(K^+ \to \pi^- e^+ e^+) \leq 2 \times 10^{-12}\,, 
\quad \text{BR}(K^+ \to \pi^- \mu^+ \mu^+) \leq 0.4 \times 10^{-12}\,. 
\end{eqnarray}

\begin{figure}[t!]
\begin{center}
\includegraphics[width=.85\textwidth]{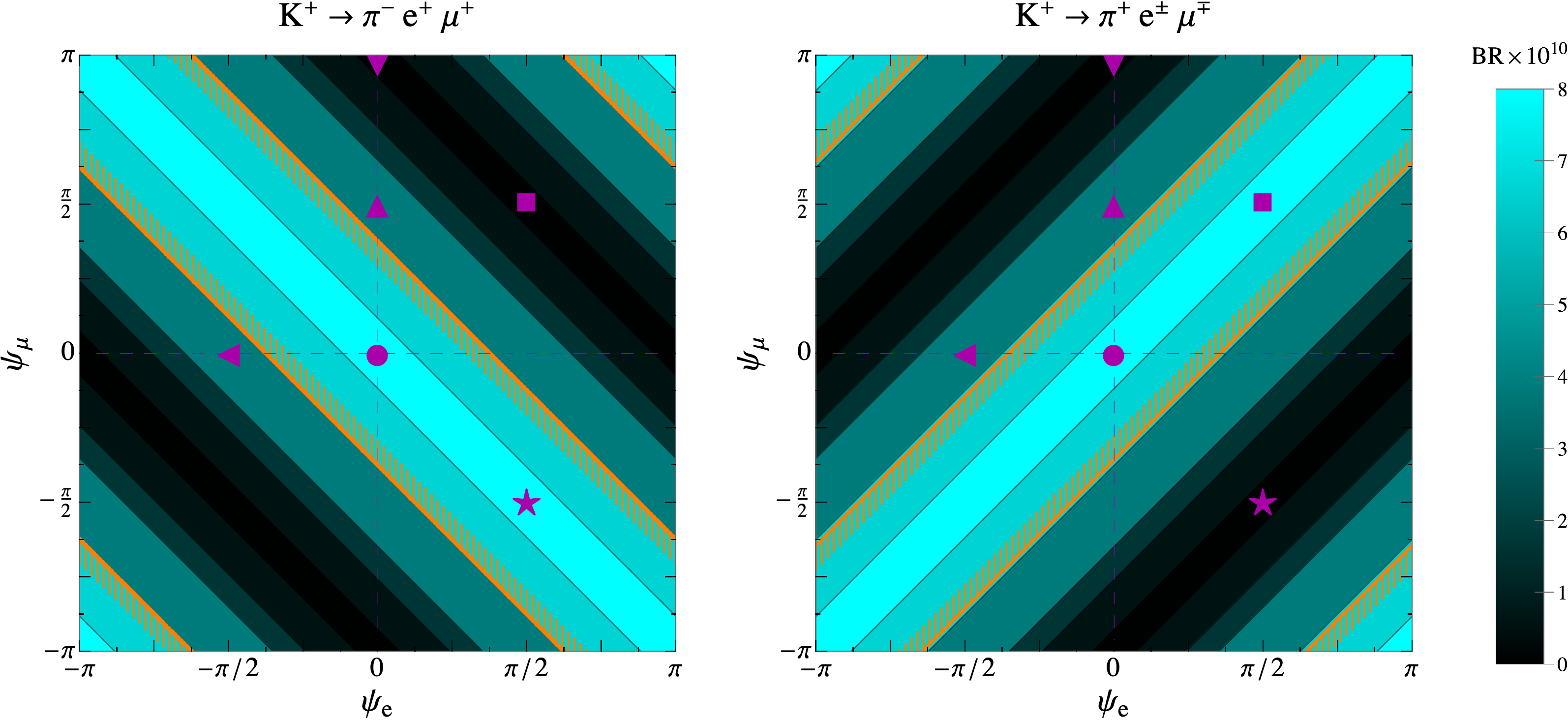}
\caption{Branching ratios of the LNV $K^+\to \pi^- e^+ \mu^+$ decay
  (left) and of the LNC $K^+ \to \pi^+ e^\pm \mu^\mp$ decay (right),
  represented via coloured isosurfaces on the $\psi_e - \psi_\mu$ plane.  
  In each panel, the orange lines denote the corresponding current
  experimental bounds (cf. Eqs.~(\ref{bounds:current},\ref{bounds:current2})). 
  The different superimposed symbols correspond to the illustrative
  points used in the discussion.}\label{plotsBRemu} 
\end{center}
\end{figure}

\begin{figure}[t!]
\begin{center}
\includegraphics[width=.55\textwidth]{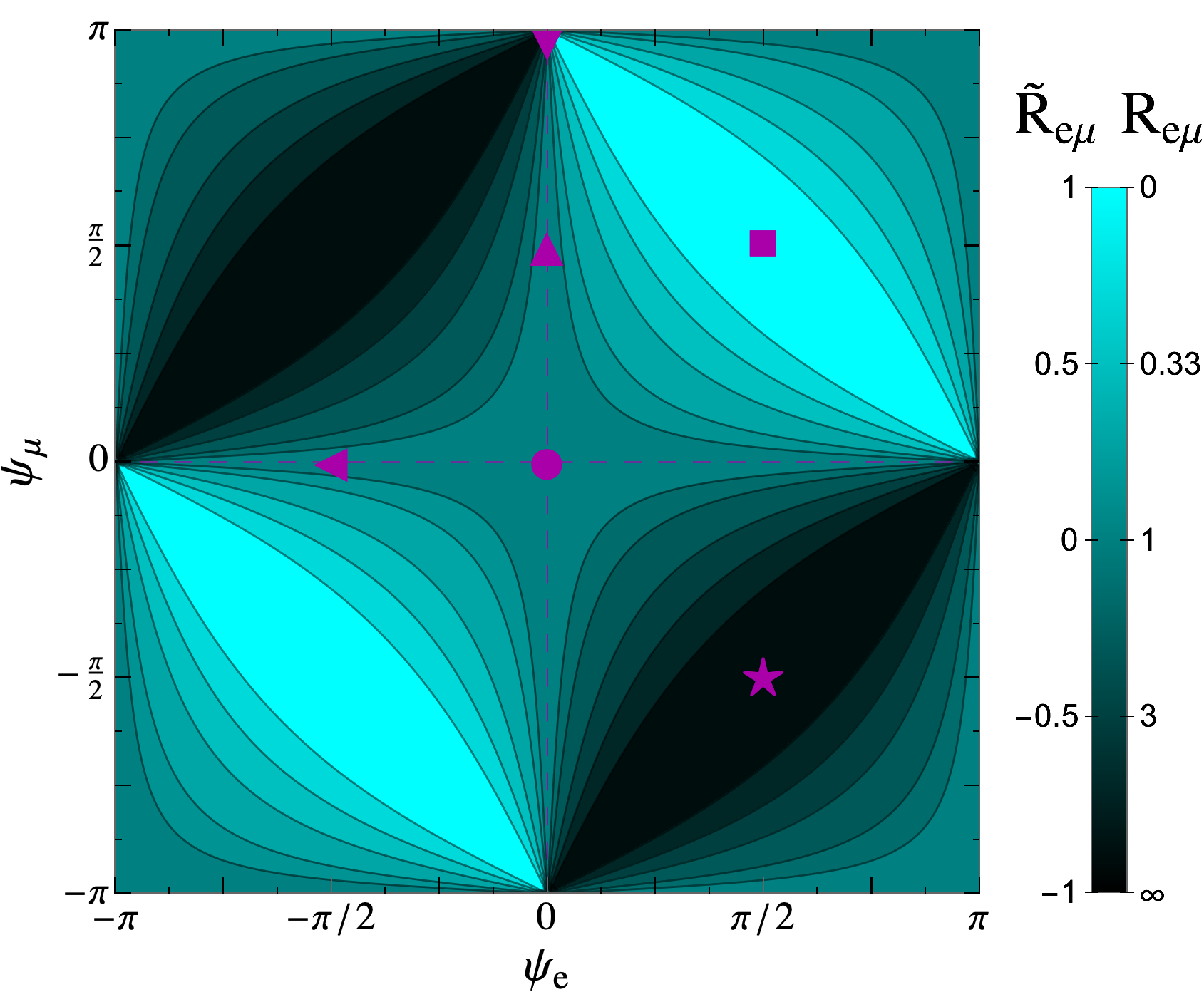}
\caption{Ratios $R_{e\mu}$  and $\widetilde R_{e \mu}$  
  as defined in Eqs.~(\ref{ratio:simple},~\ref{Rtilde}) 
  of the LNV and LNC processes $K\to\pi e \mu$.
  The coloured isosurfaces on the $\psi_e - \psi_\mu$ plane denote
  regimes for $R_{e \mu}$ and  
  $\widetilde R_{e\mu}$. Symbols as in Fig.~\ref{plotsBRemu}.
}\label{plotRprimeemu}
\end{center}
\end{figure}

We begin by displaying in Fig.~\ref{plotsBRemu} 
the predictions for each of the observables, 
BR($K^+ \to \pi^- e^+\mu^+$) and 
BR($K^+ \to \pi^+ e^\pm \mu^\mp$),
on the parameter space spanned by the two relative phases, 
$\psi_e$ and $\psi_\mu$. (We recall
that $\psi_\alpha \equiv \phi_{\alpha 4} -\phi_{\alpha 5}$, with $\phi_{\alpha i}$ 
defined in Eq.~(\ref{angle-phase})). 
The coloured isosurfaces denote regimes of the corresponding BRs,
going from $\mathcal{O}(10^{-10})$ (cyan) to nearly 0 (black); the
orange lines denote the corresponding curent bounds, see Eqs. (\ref{bounds:current},\ref{bounds:current2}). Notice that on the right panel, we choose to display only the most conservative of the bounds in 
Eq.~(\ref{bounds:current}) (to avoid overloading the plot).
For the sake of illustration, 
leading to the numerical results
of Fig.~\ref{plotsBRemu}, we have taken, 
$M=350$~MeV with $\Delta M \approx 0$ and    
$|U_{\alpha i}|=\mathcal{O}(10^{-5})$ ($\alpha=e, \mu$ and $i=4,5$). 
The associated ratios, $R_{e\mu}$ and
$\widetilde R_{e\mu}$ (see Eqs.~(\ref{ratio:simple}, 
\ref{Rtilde})) are displayed in Fig.~\ref{plotRprimeemu}, in which the
coulored isosurfaces reflect regimes of $\widetilde R_{e\mu}$ 
($R_{e\mu}$) ranging from -1 to 1 ($\infty$ to 0).
For completeness, let us notice that the prediction obtained in the case
in which the SM is extended by only one sterile fermion (i.e. $N=1$, 
with $m_4=350$ MeV and $|U_{e4}|=|U_{\mu 4}|\simeq 10^{-5}$), is
BR~$=1.8\times 10^{-10}$ for both LNC and LNV modes, 
leading to $R_{e\mu} = 1$.
    
Both panels of Fig.~\ref{plotsBRemu} recover and 
generalize the limiting case 
$\psi_\alpha=\psi_\beta$ considered in 
Fig.~\ref{plotRonephase}, now for the specific case of $K\to\pi e \mu$
decays. They also clearly show how the presence of (at least) two
sterile states can lead to strong deviations from the minimal
extension by a single heavy fermion: for identical values of sterile
masses and mixing angles, the variation of the relative phases 
$\psi_e$, $\psi_\mu$ can lead to very different predictions 
for the BRs. 

\noindent
For the left panel, the impact on 
BR($K^+ \to \pi^- e^+\mu^+$) is particularly evident 
along the direction defined by $\psi_e=\psi_\mu$, as one can go
from BRs already in conflict with experimental  bounds for 
$\psi_e=\psi_\mu=0$ to a situation of 
BR($K^+ \to \pi^- e^+ \mu^+$)$\approx 0$, due to a maximal destructive
interference of the heavy fermion contributions occurring for instance for 
$\psi_e=\psi_\mu=\pm \pi/2$. 

\noindent
Along the $\psi_e=\psi_\mu$ direction, 
no effect is manifest for the LNC modes; as seen from Eq.~(\ref{GamamLNC}), 
these depend on the combination $\psi_\alpha - \psi_\beta$, in 
which the Majorana phases $\varphi_{4,5}$ cancel out. 
This can be verified
for the LNC $K^+ \to \pi^+ e^\pm \mu^\mp$ decays, displayed on the
right panel of Fig.~\ref{plotsBRemu}, in which 
isosurfaces of constant BR are ``parallel''
to the $\psi_e=\psi_\mu$ direction. 
However, any departure  from the latter 
opens the door to interference effects, which are all the more visible
for $\psi_e=-\psi_\mu$, the direction along which the
interference effects due to both Majorana and Dirac phases 
are maximal for the LNC decays.  
As occurred for the LNV case (left panel), 
it is visible on the right panel that one goes
from BR($K^+ \to \pi^+e^\pm \mu^\mp$) already excluded by  curent
data to cases of maximal destructive interference. 

\noindent 
Due to the ``orthogonal''  dependence on the relative phases 
($\psi_e\pm\psi_\mu$, respectively for the LNV and LNC decays),
the cancellation leading to the extreme case of vanishing 
LNC amplitudes corresponds in some cases to maximal values for the LNV 
ones (and vice-versa): specific examples are 
$(\psi_e,\psi_\mu)=(\pi/2,\pi/2)$ and 
$(\psi_e,\psi_\mu)=(\pi/2,-\pi/2)$, depicted in 
Fig.~\ref{plotsBRemu} by {\footnotesize $\blacksquare$} and 
{\footnotesize $\bigstar$}.

The results of Fig.~\ref{plotsBRemu} clearly illustrate the 
role of the constructive/destructive interferences regarding 
the potential observation of each transition, and strongly suggest 
that any conclusion 
regarding the contribution of sterile fermions to LNV semileptonic meson
decays must be accompanied by the study of the corresponding (flavour violating)
LNC mode. Even if a combination of phases leads to an experimentally
``blind spot''  in which the LNV BR lies beyond sensitivity due to destructive
interference effects, the same interference might be constructive for
the corresponding LNC mode, which
could then be associated with a large BR (and vice-versa).  

The LNV decay modes leading to same-flavoured final
state\footnote{We do not show the
  numerical results for these channels, although we 
  have verified that they also follow Eq.~\eqref{GamamLNV}.} 
leptons (i.e. $\mu^\pm \mu^\pm$ and $e^\pm e^\pm$)  are also sensitive 
to the relative $CP$ violating phases. 
The interference for the same flavour final
states is maximally constructive (destructive) along the 
$\psi_\ell=0~ (\pm \pi/2)$ directions 
(as can be seen 
from Eq.~(\ref{GamamLNV})). 
For instance, some of 
the regions around $(\psi_e, \psi_\mu) = (0, \pm \pi/2)$ and
$(\pm \pi/2, 0)$, that are associated with maximal/minimal values of    
BR($K^+ \to \pi^- e^+ e^+ (\mu^+ \mu^+)$),  could lead 
to BRs for the different-flavour final states 
(both LNV and LNC) well beyond experimental sensitivity (and conversely).

\bigskip
The above discussion and the inferred conclusions become
even more straightforward when cast in terms of the
ratios $R_{e \mu}$ and $\widetilde R_{e\mu}$, as shown in 
Fig.~\ref{plotRprimeemu}. 
Let us then consider several possible experimental scenarios, for
instance in association with a future measurement of the LNV and/or LNC
kaon decays by NA62 (see Eqs.~(\ref{NA62:future}) for the future 
experimental sensitivity).

\noindent 
If (within experimental precision) 
both measurements are approximately compatible with $R_{e \mu}\sim 
1$, then interpreting an observation in terms of processes
mediated by an on-shell sterile neutrino either calls upon scenarios
with a single state, or then extensions via several states, either  with
sizeable mass differences, or then close in mass but with very
specific configurations of the $CP$ phases   (all cases leading to 
$R_{e \mu}=1$ and consequently to $\widetilde R_{e\mu}=0$).

\noindent
If the NA62 results translate into a ratio $R_{e\mu}<1$, then an
interpretation in terms of a SM extension via sterile states would
imply that one is in the presence of at least two additional
fermions,
both with masses within the region allowing for on-shell production
(approximately between $\sim$ 0.1 GeV and 0.5 GeV); more importantly
it clearly suggests that the sterile fermions are on 
the regime leading to significant interference effects (see
Eq.~(\ref{conditions})). Both measurements, i.e.  
$K^+ \to \pi^-e^+\mu^+$ and $K^+ \to \pi^+ e^- \mu^+$
could then be translated into corresponding regions in the 
$\psi_e - \psi_\mu$ plane. 
It is worth emphasising that, as can be inferred from
Eqs.~(\ref{bounds:current} - \ref{NA62:future}), measurements can in
principle lead to $R_{e\mu}$ as small as $\sim 10^{-2}$. 
As already suggested from the general discussion of
Section~\ref{results} (cf. Fig.~\ref{plotRonephase}, in which the
limiting case $\psi_e=\psi_\mu$ had been considered), such sizeable deviations
from $R_{e \mu}\sim 1$ strongly point towards a scenario of 
$\Delta M/\Gamma_N \ll 1$, and to very strong destructive interference
effects in the LNV amplitude. This corresponds to $\psi_e \sim
\psi_\mu\sim \pm \pi/2$, as can be verified in Fig.~\ref{plotRprimeemu}.

A possible measurement of the LNV same-flavoured lepton final state
BRs would then be highly complementary, and would allow to put
our working hypothesis  to the test (i.e. the presence of at least two 
Majorana fermions, with very close masses and 
non-vanishing relative $CP$ phases). If the two
additional observations\footnote{The observation of the $K^+\to \pi^-
  \mu^+ \mu^+$, if interpreted under the current hypotheses, would
  further constrain the masses of the sterile neutrinos, $m_i\in \sim
  [250,390]$ MeV, see~\cite{Abada:2017jjx}.\label{footnotemumu}}  
are found to be compatible with the model's
predictions (for the hinted regimes of the phases), this would
naturally strenghten the underlying assumptions. 
The symbols $\blacktriangle$ 
and $\blacktriangleleft$ respectively corresponding to
$(\psi_e,\psi_\mu) \approx (0, \pi/2)$ and  $(\psi_e,\psi_\mu) \approx
(- \pi/2,0)$,  
illustrate 
this: partial destructive interferences in 
$K^+\to \pi^\mp e^+\mu^\pm$ would lead to $R_{e \mu}\sim 1$,
potentially observable $K^+ \to \pi^- e^+ e^+ (\mu^+ \mu^+)$, but a
maximal destructive interference in 
$K^+ \to \pi^- \mu^+ \mu^+ (e^+ e^+)$ -- which would then likely lie beyond
experimental reach. 
 
\noindent
If neither LNC and LNV modes are observed,
the role of the LNV decays into same-flavoured final state
leptons is also very important: a possible measurement at NA62 could
still point towards a SM extension via at least two sterile neutrinos, 
whose relative phases account for a full destructive interference in
the different-flavour final states, and for a constructive one in the 
same-flavoured LNV decay modes. 
For instance, this could occur for   
$(\psi_e,\psi_\mu) \approx (0, \pi)$, denoted by $\blacktriangledown$
on Figs.~\ref{plotsBRemu} and~\ref{plotRprimeemu}.

\noindent
Likewise, following the same reasoning
as above, one can also have $R_{e\mu}>1$, which would be 
a consequence of an important
destructive interference on the LNC modes.

All the discussion here held in terms of $R_{e\mu}$ can also be
translated in terms of $\widetilde R_{e\mu}$ (defined in
Eq.~(\ref{Rtilde})), which has the advantage of lying in the range 
$[-1,1]$, depending on the effect of the interferences.
Figure~\ref{plotRprimeemu} summarises the
information conveyed by both panels of Fig.~\ref{plotsBRemu}, and
suggests how  an experimental observation of the LNV and/or LNC
channels, if interpreted under the current hypothesis, could help
determining the relative $CP$ phases of the sterile fermions. \\ 

Although Fig.~\ref{plotRprimeemu} corresponds to an analysis of the
kaon sector,  
 given that interference effects on any meson semileptonic LNV and LNC decay
lead to the same behaviour in terms of the ratios
 $\widetilde R_{\alpha \beta}$ or $R_{\alpha \beta}$ 
(for $\alpha \neq \beta$), it is important to notice that 
Fig.~\ref{plotRprimeemu} in fact illustrates in a generic way the 
impact of the $CP$ phases that might be
present in extensions of the SM by at least two sterile fermions.

\bigskip
Finally, we emphasise that a crucial outcome of our analysis is that 
a non-observation of the LNV (or LNC) mode
should not be directly
translated into more stringent bounds for the sterile mixing angles;
it could instead suggest that more than one sterile fermion is present, and
interference effects could lead to deviations from the case of a
single state. A particularly important consequence concerns finals
states with different flavoured charged leptons: if {\it
  only} an LNC mode is observed (even if the associated LNV rate is
expected to be within experimental reach), this does not imply that the heavy
mediator is a Dirac particle; on the contrary, the processes can be
mediated by two Majorana states whose relative phases are at the
origin of a maximally destructive interference in the LNV modes (we
refer to the case illustrated by {\footnotesize$\blacksquare$} on
Figs.~\ref{plotsBRemu} and~\ref{plotRprimeemu}. 
(An extreme case would correspond to having two nearly degenerate
Majorana states  combining to form a pseudo-Dirac fermion leading to
cancellation of all LNV channels, and thus to $R_{\alpha \beta}=0$, as
occurs in low scale seesaw mechanisms with approximate lepton number
conservation.)

\section{Conclusions} \label{concs} 

In this work we have studied for the first time 
the impact of constructive and
destructive interference effects on the contributions of sterile
Majorana fermions (such as RH neutrinos) to the 
decay rates of lepton number conserving and
lepton number violating semileptonic  decays.

Previous studies of LNV meson decays 
have focused on the resonant enhancement of the corresponding 
widths due to the presence of an on-shell Majorana fermion; 
the additional sterile state can also mediate LNC semileptonic
decays; the decay rates into same-sign and
opposite-sign dileptons are expected to be of the same order 
(i.e., the ratio $R_{\ell_\alpha \ell_\beta}=1$). 

We have extended these studies, now taking into account the presence of at
least two sterile neutrinos (with masses allowing  
for on-shell production), and associated $CP$ violating Dirac and
Majorana phases. Provided the mass difference of the new states 
is small enough ($\Delta M /\Gamma_N \leq 1$), we have shown that 
the $CP$ violating phases
can lead to constructive/destructive interference effects in the
BRs of the decays, reflected in deviations from  
$R_{\alpha \beta}=1$. 
We also highlighted that a non-observation of the LNV (or LNC) mode
should not be directly
translated into more stringent bounds for the sterile mixing angles;
it could instead suggest that more than one state is present.
Furthermore, observing only lepton number 
conserving modes (and no LNV transitions) 
does not rule out that the mediators are indeed of Majorana nature.

Our analysis strongly motivates a re-interpreation of the curent
negative search results for LNV and LNC semileptonic meson decays in
terms of more than one sterile neutrino; this calls upon a thorough
study in terms of an enlarged parameter space. This is the object of
an ongoing work~\cite{in-progress}.

We have illustrated the interference effects for several semileptonic 
kaon decays: the LNV $K^\pm \to\ell_\alpha^\pm\ell_\beta^\pm \pi^\mp$
($\alpha,\beta = e \text{ and/or } \mu$),
and the corresponding lepton number conserving processes,  
$K^\pm \to\ell_\alpha^\pm\ell_\beta^\mp \pi^\pm$ ($\alpha \neq \beta$),
all presently searched for in NA62. We have discussed
the possible $R_{e \mu}$ scenarios arising from a future
observation of the corresponding decays, and emphasised the relevance
of taking into account {\it all} four decay modes. Not only this
allows to refine the information on the relative phases, but offers
independent means to probe the underlying SM extension via 
sterile fermions.

A future observation of such processes - if interpreted under the
proposed paradigm of at least two sterile fermions - opens a
unique window onto the heavy sector parameter space: firstly their
average sterile mass is constrained to lie on the resonant production
interval;  secondly, 
one can identify ranges of variation for the
different couplings (in the case of kaon decays, $|U_{ei}|$  and 
$|U_{\mu i}|$); finally, information can also be inferred on the
relative $CP$ violating phases (even if not disentangling the Dirac
from the Majorana ones).
This information can be ultimately used to probe if the
heavy Majorana fermions responsible for the observed LNV
and LNC transitions are an integral part of a given
mechanism of neutrino mass generation.

\section*{Acknowledgements}
We acknowledge support within the framework of the
European Union’s Horizon 2020 research and innovation programme
under the Marie Sklodowska-Curie grant agreements No 690575 and No
674896.

\end{document}